\documentclass[aps,prd,nofootinbib,amsmath,twocolumn]
{revtex4}
\usepackage{amssymb,esvect,amsmath,graphicx,latexsym,amsthm,slashed,eso-pic,amsfonts,hyperref}
\usepackage[final]{pdfpages}
\newcommand{\be}{\begin{equation}}
\newcommand{\ee}{\end{equation}}
\newcommand{\nl}{\nonumber \\}

\usepackage{verbatim}

\newcommand{\TeV}{\text{ TeV}}
\newcommand{\GeV}{\text{ GeV}}
\newcommand{\MeV}{\text{ MeV}}

\newcommand{\eV}{\text{ eV}}

\newcommand{\x}{\chi}
\newcommand{\xp}{\chi^\prime}

\def\lsim{\mathrel{\raise.3ex\hbox{$<$\kern-.75em\lower1ex\hbox{$\sim$}}}}
\def\gsim{\mathrel{\raise.3ex\hbox{$>$\kern-.75em\lower1ex\hbox{$\sim$}}}}

\newcommand{\order}[1]{\mathcal{O}{(#1)}}

\begin{document}

\hspace{13cm} \parbox{5cm}{SLAC-PUB-16953}~\\

\hspace{13cm}

\title{WIMPs with GUTs: Dark Matter Coannihilation with a Lighter Species}
\author{Asher Berlin}

\affiliation{SLAC National Accelerator Laboratory, 2575 Sand Hill Road, Menlo Park, CA, 94025, USA}

\date{\today}

\begin{abstract}

We propose a new thermal freeze-out mechanism for ultra-heavy dark matter. Dark matter coannihilates with a lighter unstable species, leading to an annihilation rate that is exponentially enhanced relative to standard WIMPs. This scenario destabilizes any potential dark matter candidate. In order to remain consistent with astrophysical observations, our proposal necessitates very long-lived states, motivating striking phenomenology associated with the late decays of ultra-heavy dark matter, potentially as massive as the scale of grand unified theories, $M_\text{GUT} \sim 10^{16}$ GeV.

\end{abstract}

\maketitle

The Weakly Interacting Massive Particle (WIMP) paradigm has motivated searches for dark matter (DM) particles with weak-scale masses and interactions with the Standard Model (SM). In this scenario, the large thermal number density of DM is depleted through $2 \to 2$ processes that eventually freeze out of chemical equilibrium once the associated rate drops below the expansion rate of the universe. The observation that weak-scale masses and couplings give rise to an abundance of WIMPs that is in agreement with the observed DM energy density is often referred to as the ``WIMP miracle." This narrative provides a useful benchmark that has guided the experimental community for decades. However, in spite of their allure, WIMPs have alluded detection to date; the Large Hadron Collider (LHC) has not yet observed definite signs of new physics~\cite{Aad:2015zva,Khachatryan:2014rra}, and limits from null results of direct detection experiments have grown at an exponential rate~\cite{Tan:2016zwf,Akerib:2015rjg,Akerib:2016vxi,Agnese:2015nto}.

One plausible explanation for the absence of discovery is that DM has a mass that is much larger than the electroweak scale. For $m_{_\text{DM}} \gg \order{100} \GeV$, the LHC center of mass energy is insufficient to create a significant number of DM particles in proton collisions, and the suppressed number density, $n_{_\text{DM}} \propto 1 / m_{_\text{DM}}$, limits the ability of direct detection or astrophysical searches. The situation is exacerbated if DM is additionally a SM singlet. In this case, DM resides in a hidden sector (HS) that is populated independently following post-inflation reheating~\cite{Feng:2008mu,Hardy:2017wkr,Adshead:2016xxj}. The DM abundance is then depleted through annihilations to lighter HS states that possess feeble couplings with SM particles~\cite{Pospelov:2007mp}.

Within the WIMP framework, perturbative unitarity of the theory, supplemented with astrophysical data, limits the DM mass to be $m_{_\text{DM}} \lesssim \order{10^5} \GeV$~\cite{Griest:1989wd}. However, it has long been appreciated that this bound can be circumvented by invoking a non-standard cosmological history~\cite{Fornengo:2002db,Kane:2015jia,Hooper:2013nia,Randall:2015xza,Reece:2015lch,Lyth:1995ka,Davoudiasl:2015vba,Cohen:2008nb,Berlin:2016vnh,Berlin:2016gtr}. Large injections of entropy into the SM bath dilute the current abundance of any relic species by effectively increasing the present age of the universe. 
For example, if a particle of comparable mass decays  after DM freezes out
with $\order{1}$ couplings, then
\be
\label{eq:massub}
m_{_\text{DM}} \sim \order{10^{-4}} ~ \frac{T_\text{eq} \, m_\text{pl}}{T_{RH}}  \lesssim 10^{8} \GeV
~.
\ee
In Eq.~(\ref{eq:massub}),  $T_\text{RH}$ and $T_\text{eq} \simeq 0.8 \eV$ are the temperatures of the SM bath after the decay and at matter-radiation equality, respectively. The inequality is a result of demanding that $T_\text{RH} \gtrsim 10 \MeV$, in agreement with the successful predictions of Big Bang nucleosynthesis (BBN). Completely saturating perturbative unitarity increases the above upper bound by $\order{10}$.

In this Letter, we propose a new thermal freeze-out mechanism in which $m_{_\text{DM}} \gg 10^{10} \GeV$. With this philosophy in mind, we will not focus heavily on an explicit model, but will discuss a simplified Lagrangian that exhibits the required dynamical structure. DM coannihilates with a lighter unstable HS species, leading to an annihilation rate that is exponentially enhanced relative to standard WIMPs. DM chemically decouples once the number density of the \emph{lighter} species is sufficiently diluted by Hubble expansion,  effectively delaying freeze-out.   In the simplest constructions, this scenario destabilizes any potential DM candidate. To remain consistent with astrophysical observations, our proposal necessitates very long-lived states, motivating striking phenomenology associated with the late decays of ultra-heavy DM, potentially as massive as $M_\text{GUT} \sim \order{10^{16}} \GeV$. Non-thermal mechanisms for producing ultra-heavy DM have been studied previously, e.g., in the context of inflation, gravitational production, or large entropy injections~\cite{Kolb:1998ki,Chung:1998rq,Chung:2001cb,Feldstein:2013uha,Harigaya:2014waa,Hui:1998dc,Bramante:2017obj,Harigaya:2016vda}. To the best of our knowledge, the processes studied in this work represent the first investigation of superheavy DM that is generated from \emph{thermal} freeze-out in a sector that has comparable energy density to the SM bath at early times.

Coannihilation between two different DM species has long been investigated as a possible contributor to the freeze-out process~\cite{Griest:1990kh,Edsjo:1997bg}. In the standard example, DM, denoted by $\x$, coannihilates with a slightly heavier species, $\xp$, into a pair of SM particles, $\x \, \xp \to \text{SM} ~ \text{SM}$. $\x$ and $\xp$ are often assumed to both be charged under a $\mathbb{Z}_2$ symmetry, while $m_\x < m_{\xp}$ ensures the cosmological stability of $\x$. To keep our discussion pedagogical, we will momentarily consider the modified coannihilation process in which $m_\x \gg m_{\xp}$. As we will show, this does not constitute a viable freeze-out paradigm, but is useful in illustrating why related scenarios may have been overlooked in the past. The thermal DM population departs from chemical equilibrium once the coannihilation rate becomes comparable to the Hubble parameter, $H$,
\be
\label{eq:fo1}
n_{\xp} ~ \langle \sigma v \rangle \sim H
~,
\ee
where $n_{\xp}$ is the number density of $\xp$. In order for $\x$ to be a realistic candidate for the universe's cold DM, we will demand that this occurs at temperatures below its mass, i.e., $x_f \equiv m_\x / T_\text{FO} > 1$.  On the other hand, $m_\x \gg m_{\xp}$ implies that $\xp$ is relativistic with a corresponding number density at freeze-out of $n_{\xp} \sim (m_\x / x_f)^3$. 

Parametrizing the thermally-averaged coannihilation cross section as $\langle \sigma v \rangle \equiv \alpha_\x^2 / m_\x^2$, Eq.~(\ref{eq:fo1}) can be rewritten as
\be
\label{eq:fo2}
m_\x \sim \order{10^{-2}} ~ (\alpha_\x^2 / x_f) ~ m_\text{pl}
~.
\ee
The same process that governs freeze-out also allows $\x$ to decay, e.g., $\x \to \xp \text{ SM SM}$, with a width that scales as $\Gamma_\x \sim \alpha_\x^2 \, m_\x$. Stability of $\x$ demands that $\Gamma_\x \lesssim H_0$, where $H_0$ is the Hubble parameter today. Along with Eq.~(\ref{eq:fo2}), this implies
\be
\label{eq:fo3}
m_\x \ll  \order{1} \eV
~.
\ee
Eq.~(\ref{eq:fo3}) obviously contradicts the requirement that $\x$ is non-relativistic at the time of matter-radiation equality.

\begin{figure}[t]
\hspace{-0.5cm}
\includegraphics[width=0.45\textwidth]{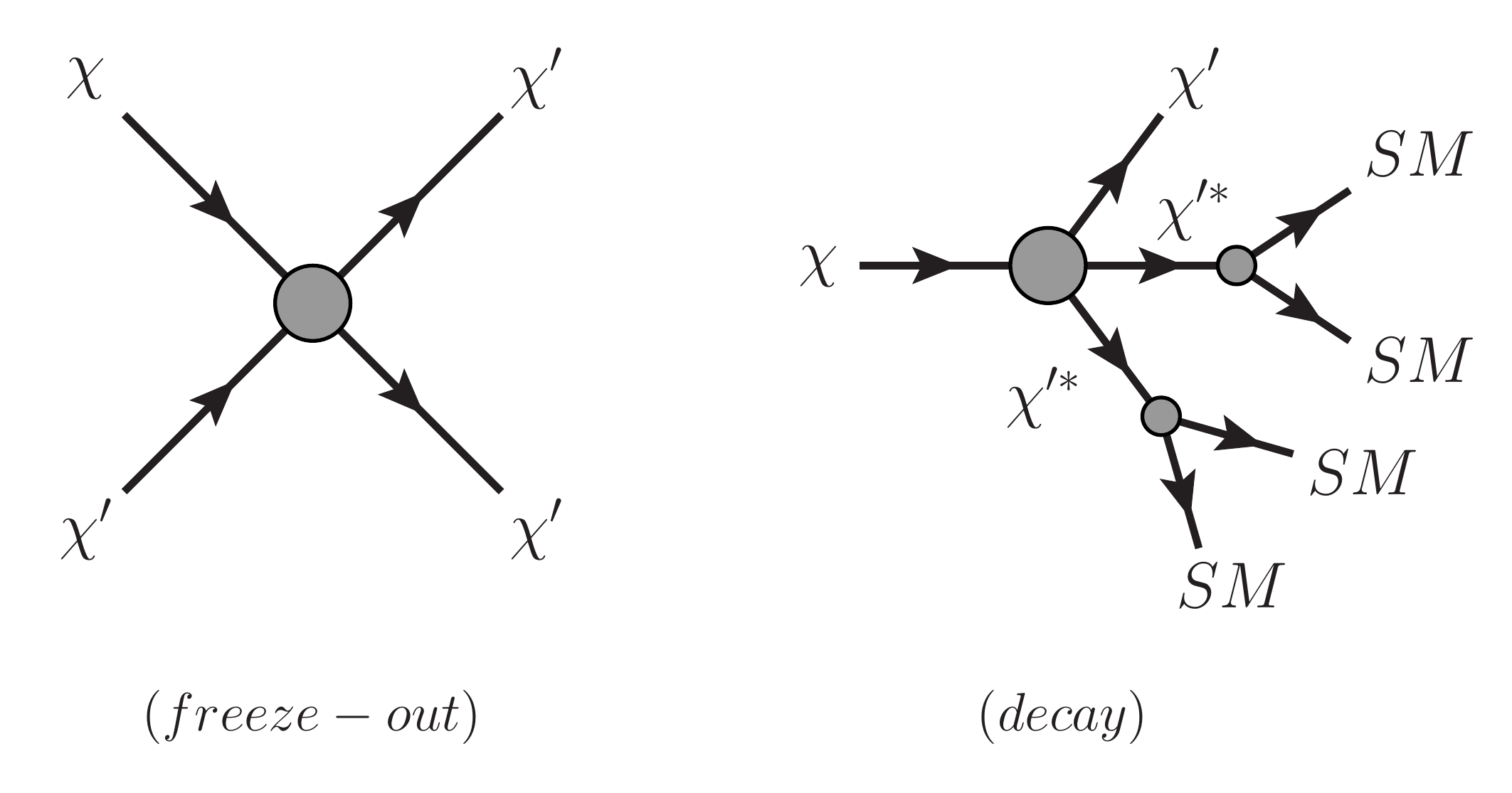} \hspace{-0.5cm}
\caption{Representative Feynman diagrams for the freeze-out (left) and decay (right) of dark matter, $\x$. The lighter state in the hidden sector, $\xp$, is assumed to couple to a pair of Standard Model particles. }
\label{fig:FD}
\end{figure}

We will aim to modify the above scenario such that $\x$ is long-lived and is a viable DM candidate.\footnote{One possible modification involves $3 \to 2$ processes where both DM and a lighter species are in the initial state. This has been explored in the context of light DM in Refs.~\cite{Cline:2017tka,Dey:2016qgf}.} Let us assume that a HS is thermally populated during the period of reheating that follows inflation and possesses very feeble interactions with the SM~\cite{Allahverdi:2010xz,Adshead:2016xxj,Hardy:2017wkr}. The hidden and visible sectors are taken to be kinetically decoupled, such that each tracks a distinct thermal distribution governed by the temperatures $T_h$ and $T$, respectively. In general, $T_h \neq T$, and we define the corresponding ratio immediately after inflation as
\be
\xi_\text{inf} \equiv T_h / T ~ \big|_\text{inf}
~.
\ee
The time-evolution of $\xi \equiv T_h / T$ can be derived from its initial value, $\xi_\text{inf}$, and the conservation of comoving entropy density in the hidden and visible sectors. For concreteness, we will focus on $\xi_\text{inf} \sim \order{1}$. DM, denoted as $\x$, possesses non-negligible interactions with a lighter unstable HS species, $\xp$, that is nearby in mass. Without specifying the spin or specific form of the interactions, we will assume that $\x$ coannihilates with $\xp$ into a pair of $\xp$, through the process $\x \xp \to \xp \xp$. We write the corresponding thermally-averaged cross section as
\be
\label{eq:sigmav}
\langle \sigma v \rangle \equiv \frac{\alpha_\x^2}{m_\x^2} \quad (\x \xp \to \xp \xp)
~,
\ee
where $\alpha_\x$ is an effective coupling responsible for coannihilation. 
An implicit $\mathbb{Z}_3$ symmetry acting on $\xp$ forbids mass-mixing between $\x$ and $\xp$. This process is depicted in the left diagram of Fig.~\ref{fig:FD}. A toy model for this scenario will be presented towards the end of this work.

If $\xp$ is the lightest state in the HS, it is naturally long-lived, since it can only decay into visible sector final states,
\be
\label{eq:decay1}
\Gamma_{\xp} (\xp \to \text{SM SM}) \equiv \epsilon^2 ~ m_{\xp}
~.
\ee
We will demand that $\epsilon \ll 1$, which guarantees that the hidden and visible sectors remain thermally decoupled throughout the freeze-out of $\x$. Although we have written Eq.~(\ref{eq:decay1}) such that $\xp$ decays into pairs of SM particles, we will more generally consider the scenario where $\xp$ decays into any pair of particles that are thermally coupled to the photon plasma at early times. 

The cosmological evolution of $\x$ is governed by the corresponding Boltzmann equation, 
\begin{align}
\label{eq:boltz1}
&\dot{n}_\x + 3 H n_\x = - \langle \sigma v \rangle ~ \Big(n_\x - \frac{n_{\xp}}{n_{\xp}^\text{eq}} \, n_\x^\text{eq} \Big) ~ n_{\xp}
~,
\end{align}
where $n^\text{eq}$ denotes the equilibrium number density. We will assume that $\xp$ decays after the freeze-out of $\x$. In this case, the comoving entropy densities of the hidden and visible sectors are separately conserved during the freeze-out process. Hence, the above form can be rewritten in terms of the yield, $Y \equiv n / s$, and the dimensionless parameter $x \equiv m_\x / T$, 
\be
\label{eq:boltz2}
\frac{d Y_\x}{d x} = - \, \frac{s \langle \sigma v \rangle}{H x} ~ \Big( Y_\x - \frac{Y_{\xp}}{Y_{\xp}^\text{eq}} \, Y_\x^\text{eq} \Big) ~ Y_{\xp}
~,
\ee
where $s$ is the entropy density of the visible sector. Eq.~(\ref{eq:boltz2}) can be conveniently recast as
\be
\frac{x \, (d Y_\x / d x)}{Y_\x^\text{eq}} = - \, \frac{( n_{\xp}^2 / n_{\xp}^\text{eq} ) \, \langle \sigma v \rangle}{H} ~ \bigg( \frac{Y_\x ~ Y_{\xp}^\text{eq}}{ Y_{\xp} ~ Y_\x^\text{eq}} - 1 \bigg)
~,
\ee
which implies that $\x$ is unable to maintain chemical equilibrium once
\be
\label{eq:fo4}
( n_{\xp}^2 / n_{\xp}^\text{eq} ) ~ \langle \sigma v \rangle \sim H
~.
\ee

In order to estimate which values of $m_\x$ lead to an adequate DM abundance, we solve Eq.~(\ref{eq:fo4}) for $n_{\x}^\text{eq}$ and equate the DM and radiation energy density at $T_\text{eq} \simeq 0.8 \eV$. If $\xp$ remains in chemical equilibrium while $\x$ freezes out, we can set $n_{\xp} = n_{\xp}^\text{eq}$ in Eq.~(\ref{eq:fo4}). In this case, $\x$ freezes out with the proper abundance for
\be
\label{eq:fo5}
m_\x \sim e^{(1-m_{\xp}/ m_\x) x_f / 2 \xi } \times \alpha_\x ~ (T_\text{eq} \, m_\text{pl})^{1/2} 
~,
\ee
where $\xi$ is evaluated at freeze-out and we have assumed that the comoving entropy in the visible sector is approximately conserved at all times. For $\alpha_\x \sim \order{10^{-2}}$, $\alpha_\x \, (T_\text{eq} \, m_\text{pl})^{1/2} \sim \order{1} \TeV$. Thus, the second factor in Eq.~(\ref{eq:fo5}) corresponds to the usual coincidence of scales as noted in the WIMP miracle paradigm. However, for $m_{\xp} < m_\x$, the prefactor in Eq.~(\ref{eq:fo5}) is representative of the exponential enhancement in the number density of $\xp$ target-scatterers. For perturbative values of $\alpha_\x$, this favors DM masses significantly above the electroweak scale. 

Alternatively, if $\xp$ is unable to deplete its number density at temperatures below its mass, the analogue of Eq.~(\ref{eq:fo5}) becomes
\be
\label{eq:fo6}
m_\x \sim e^{(1 + m_{\xp}/ m_\x) x_f / 2 \xi } \, \xi^{3/2} \times \alpha_\x ~ (T_\text{eq} \, m_\text{pl})^{1/2} 
~.
\ee
While still significantly greater than the electroweak scale, the size of $m_\x$ that is favored in Eq.~(\ref{eq:fo5}) is exponentially smaller than that of Eq.~(\ref{eq:fo6}). This is due to the additional Boltzmann suppression of $n_{\xp}$ when $\xp$ remains chemically coupled throughout the freeze-out of $\x$. 

The processes shown in Eqs.~(\ref{eq:sigmav}) and (\ref{eq:decay1}) allow $\x$ to deplete its number density through coannihilations with $\xp$. These same interactions unavoidably lead to its decay through $\x \to \xp \x^{\prime *} \x^{\prime *} \to \xp \text{ SM} \cdots$, as shown in the right diagram of Fig.~\ref{fig:FD}. For instance, 5-body tree-level and 3-body loop-level decays are possible, with the corresponding widths scaling as
\begin{align}
\label{eq:decay2}
& ~~~~ \Gamma_\x^\text{tree} (\x \to \xp \text{ SM SM SM SM}) 
\nl
& \sim \Gamma_\x^\text{loop} (\x \to \xp \text{ SM SM}) \sim \frac{\epsilon^4 ~ \alpha_\x^2}{192 \pi^3 \, (4 \pi)^4} ~ m_\x
~,
\end{align}
where we have assumed that each $\xp$ decays to a pair of SM particles and hence have included $192 \pi^3 \, (4 \pi)^4$ as an approximate 5-body phase-space factor 
or 3-body and loop factor. We will focus on mass hierarchies of roughly $1 < m_\x / m_{\xp} < 2$, such that the analogous decays $\x \to \xp \xp \x^{\prime *} \to \xp \xp \text{ SM SM}$ and $\x \to \xp \xp \xp$ are kinematically forbidden. 

DM decays are strongly constrained by the cosmic microwave background (CMB) regardless of the precise identity of the final state visible sector particles in Eq.~(\ref{eq:decay2}). In particular, measurements of the CMB power spectrum place a model-independent bound on the DM decay rate, $\Gamma_\x \lesssim \order{10^{-43}} \GeV$~\cite{Poulin:2016nat}. Eq.~(\ref{eq:decay2}) then implies that
\be
\label{eq:CMB}
\epsilon \lesssim \order{10^{-11}} \times \left( \frac{\alpha_\x}{10^{-3}} \right)^{-1/2} \left( \frac{m_\x}{10^{16} \GeV} \right)^{-1/4}
~,
\ee
which, from Eq.~(\ref{eq:decay1}), bounds the lifetime of $\xp$ from below. 

Therefore, $\xp$ is long-lived and naturally comes to dominate the energy density of the universe for values of $\epsilon$ that are in accord with CMB measurements. Assuming that $\xp$ is not able to deplete its abundance once it becomes non-relativistic, we find that the energy density of $\xp$ dominates over that of the SM prior to its decay for
\be
\epsilon \lesssim 10^{-4} ~ \left( \frac{m_{\xp}}{10^{16} \GeV} \right)^{1/2}
~.
\ee
Decays of $\xp$ into SM radiation increases the entropy of the visible sector bath and dilutes the abundance of the relic $\x$ population by an amount
\be
\frac{S_f}{S_i} \simeq 1.83 ~ \langle g_*^{1/3} \rangle^{3/4} ~  \frac{m_{\xp} \, Y_{\xp}}{m_\text{pl}^{1/2} \, \Gamma_{\xp}^{1/2}}
~,
\ee
where $g_*$ is the effective number of SM relativistic degrees of freedom, and the brackets denote time-averaging over the decay of $\xp$~\cite{Kolb:1990vq}. For values of $\epsilon$ that satisfy Eq.~(\ref{eq:CMB}), $S_f / S_i \gg 1$. Including this dilution for the estimate of $m_\x$ in Eq.~(\ref{eq:fo6}), we find that the abundance of $\x$ matches the observed DM energy density for 
\be
\label{eq:fo7}
m_\x \sim \order{10^{-2}} ~ e^{(1 + m_{\xp}/ m_\x) x_f / \xi } ~ \frac{\alpha_\x^2 \, \xi^6}{x_f^4} ~ \frac{T_\text{eq} \, m_\text{pl}}{T_{RH}}
~.
\ee
Above, the temperature of the SM bath after the decay of $\xp$ is approximated by
\be
T_\text{RH} \simeq \left( \frac{5}{g_* \, \pi^3}\right)^{1/4}  \left( m_\text{pl} ~ \Gamma_{\xp}\right)^{1/2}
~.
\ee
Aside from the large exponential prefactor, the parametric form in Eq.~(\ref{eq:fo7}) is nearly identical to that of Eq.~(\ref{eq:massub}). 

\begin{figure}[t]
\hspace{-0.5cm}
\includegraphics[width=0.4\textwidth]{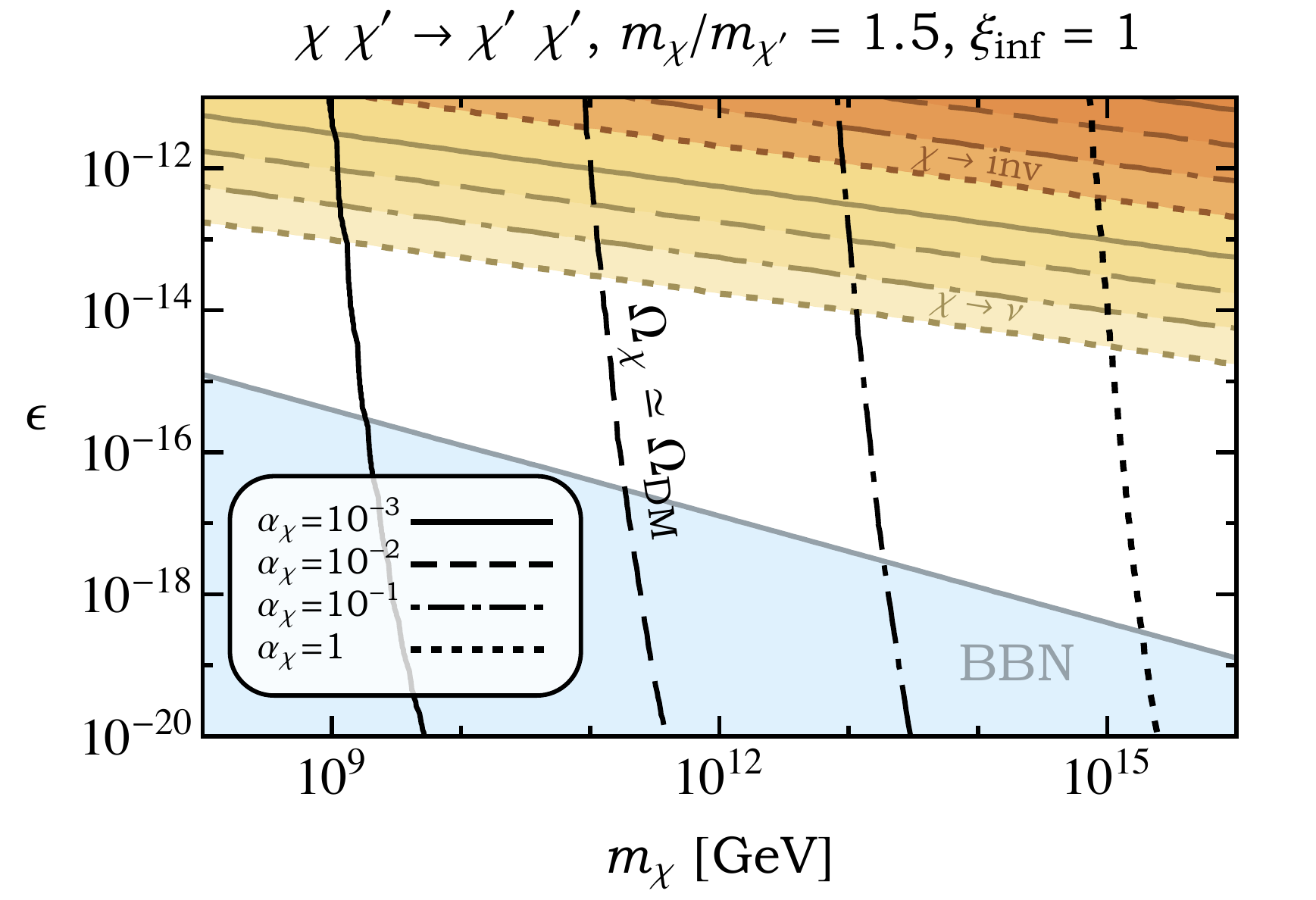} \hspace{-0.5cm}
\caption{Viable parameter space in the $\epsilon-m_\x$ plane for $\xi_\text{inf} = 1$ and $m_\x / m_{\xp} = 1.5$. $\x$ freezes out through coannihilations with a lighter unstable particle in the hidden sector, $\xp$. Along the black contours, the $\x$ abundance is equal to the measured dark matter energy density for different values of the effective coupling, $\alpha_\x$.  For sufficiently small values of $\epsilon$, the light blue region is excluded by BBN. For larger values of $\epsilon$, this model is constrained from searches for the late decays of dark matter into neutrinos by high-energy neutrino telescopes (yellow). Similar decays into SM-singlets are excluded by measurements of the CMB power spectrum (red).}
\label{fig:EpsMx}
\end{figure}

In Fig.~\ref{fig:EpsMx}, we illustrate the viable parameter space in the $\epsilon-m_\x$ plane for representative values of $\xi_\text{inf}$, $m_\x / m_{\xp}$, and $\alpha_\x$. The black contours denote regions in which the $\x$ energy density matches the observed DM abundance. In computing the relic abundance of $\x$, we have numerically solved the Boltzmann equation in Eq.~(\ref{eq:boltz1}), assuming that $\xp$ is unable to deplete its number density at temperatures below its mass. Also shown are constraints from cosmological and astrophysical probes. Preserving the successful formation of light nuclei during BBN requires that $T_\text{RH} \gtrsim 10 \MeV$, which bounds $\epsilon$ from below. 

In the case that $\xp$ (and hence $\x$) decays to SM neutrinos, ground-based neutrino telescopes place strong upper limits on $\epsilon$ for DM masses significantly above a TeV~\cite{Esmaili:2012us, Kuznetsov:2016fjt,Cohen:2016uyg}. In Ref.~\cite{Esmaili:2012us}, the non-observation of ultra-high energy neutrinos at the AMANDA, IceCube, Auger, and ANITA telescopes restricts DM lifetimes to be greater than $10^{26} - 10^{27}$ seconds for $10^4 \GeV \lesssim m_\x \lesssim 10^{16} \GeV$. For lighter masses, the solar neutrino Super-Kamiokande experiment has the greatest sensitivity, demanding $\tau_\x \gtrsim 10^{23} - 10^{25}$ seconds. More recent studies incorporating current IceCube data restrict $\tau_\x \gtrsim 10^{28}$ seconds for $m_\x \lesssim 10^7 \GeV$~\cite{Kuznetsov:2016fjt,Cohen:2016uyg}. The precise value of this limit strongly depends on the energy spectrum of the neutrinos produced in the decays of $\x$, and we will simply impose that $\tau_\x \gtrsim 10^{27}$ seconds. While a complete systematic study of the potential signals at neutrino telescopes is beyond the scope of this work, we note that in all likelihood this limit is conservative at our lower mass range and overly aggressive at larger masses. 

For $\x \to \nu + \cdots$, we show in Fig.~\ref{fig:EpsMx} the regions of parameter space that are in conflict with the null observations of neutrino telescopes. Furthermore, if $\x$ decays to invisible radiation
, we show exclusions from measurements of the CMB. This latter scenario represents the most model-independent limit on our proposed DM model, since it applies to any decay products of $\xp$. The constraints described above can be satisfied for a wide range of values for $\epsilon$. If $\alpha_\x \sim \order{1}$, $\x$ can be generated with an acceptable abundance for masses as large as $m_\x \sim 10^{15} - 10^{16} \GeV$. Larger DM masses are phenomenologically viable for $\alpha_\x  > 1$, but this may be in conflict with indirect constraints on the expansion history during post-inflation reheating~\cite{Martin:2010kz,Adshead:2010mc,Martin:2014nya,Dai:2014jja,Cook:2015vqa,Domcke:2015iaa}. Dedicated analyses of the limits derived from neutrino telescopes may shift the excluded regions of $\epsilon$. However, in Fig.~\ref{fig:EpsMx}, we expect such corrections will at most introduce additional $\order{1}$ factors since $\Gamma_\x \propto \epsilon^4$. It is enticing to note that DM masses near the scale of grand unified theories are accessible with $\order{1}$ couplings in the HS. We leave detailed model building in this regard to future work. 

A perturbative toy model can be constructed if the HS interacts with the SM through the lepton portal. We introduce a singlet DM Weyl field, $\x$, a Dirac pair of singlet right-handed neutrinos, $N$ and $N^c$, and a complex scalar, $\phi$. $N$, $N^c$, and $\phi$, as well as the SM lepton fields, $L$ and $e^c$, are charged under an unbroken global $\mathbb{Z}_3$ symmetry, as shown in Table~\ref{tab:model}. Simplifying to the case of a single generation of SM leptons, the HS Lagrangian then contains the following renormalizable interactions
\begin{align}
\label{eq:toy}
-\mathcal{L} &\supset \frac{1}{2} \, m_\x \, \x^2 + \, m_N \, N \, N^c + m_\phi^2 \,  |\phi|^2
\nl
& + \, \lambda_\x \, \x \, \left( \phi \, N^c + \phi^\dagger N \right) + \lambda_N \, \left( \phi \, N^2 + \phi^\dagger \, N^{c \, 2} \right)
\nl
& +\, y_\nu \, N \, L \, H \, + \text{h.c.}
~,
\end{align}
where 2-component Weyl and $SU(2)_L$ indices are implied, $H$ is the SM Higgs, and $m_\phi > m_\x$. In the second line above, we have demanded that interactions amongst HS particles respect the parity symmetry, $\mathbb{P}_\text{HS}  : N \leftrightarrow N^c$, $\phi \leftrightarrow \phi^\dagger$. 

\begin{table}[t]
\centering
\begin{tabular}{| c || c| c | c  | c |}
\hline
 & Spin & $SU(2)_L$ & $U(1)_Y$ & $\mathbb{Z}_3$  \\ \hline \hline
$\x$ & 1/2 &\bf{1} & 0 & 0   \\ \hline
$\phi$ & 0 & \bf{1} & 0 & +1 \\ \hline 
$N$ & 1/2 & \bf{1} & 0 & +1   \\ \hline 
$N^c$ & 1/2 & \bf{1} & 0 & -1   \\ \hline 
$L$ & 1/2 & \bf{2} & -1/2 & -1   \\ \hline
$e^c$ & 1/2 & \bf{1} & +1 & +1   \\ \hline
\end{tabular}
\caption{An example charge assignment for a lepton portal toy model. $\x$, the dark matter, freezes out through interactions with a Dirac pair of right-handed singlet neutrinos, $N$ and $N^c$, which play the role of $\xp$ discussed throughout this work. $L$ and $e^c$ are the Standard Model lepton fields. A global $\mathbb{Z}_3$ symmetry generalizes Standard Model lepton number and ensures that $\x$ is cosmologically long-lived.}
\label{tab:model}
\end{table}

In this model, $N$ and $N^c$ play the role of $\xp$ as discussed throughout this work. $\x$ freezes out with the proper DM abundance through the process $\x \, N^{(c)} \to \phi^* \to N^{(c)} \, N^{(c)}$. At later times, $N$ and $N^c$ decay to $H \, \nu$, $Z \, \nu$, and $W^\pm \ell^\mp$. 
This toy model matches onto the phenomenology discussed above through the identifications $\alpha_\x \sim \lambda_\x \, \lambda_N \, (m_\x / m_\phi)^2$ and $\epsilon \sim y_\nu$. It is technically natural to take $y_\nu \ll 1$ since non-zero values explicitly break  $\mathbb{P}_\text{HS}$ and an accidental $\mathbb{Z}_2$ under which $\x$, $N$, and $N^c$ are charged. 

In Eq.~(\ref{eq:toy}), we have neglected writing down higher-order operators that induce $\x$ decay. It is simple to see that it is technically natural for the coefficient of any such operator to be proportional to $y_\nu^2 \sim \epsilon^2 \ll 1$. Since $m_\x < m_\phi , \, 2 m_N$ these may enter in one of several ways. Operators of the form $\x ~ \mathcal{O}_\text{SM}$, $\x ~ N^{(c)} ~ \mathcal{O}_\text{SM}$, and $\x ~ (N + N^c) ~ \mathcal{O}_\text{SM}$ violate a $\mathbb{Z}_2$, $\mathbb{P}_\text{HS}$, and $\mathbb{Z}_3$ symmetry, respectively, where $\mathcal{O}_\text{SM}$ is some collection of SM fields. $\x-N^{(c)}$ mass mixing is the only process that can lead to decay rates suppressed by two powers of $y_\nu$, which is forbidden by the $\mathbb{Z}_3$. Hence, at leading order, $\Gamma_\x \sim y_\nu^4$, corresponding to the processes previously considered in Eq.~(\ref{eq:decay2}). 

Below the electroweak scale, the seesaw mechanism leads to the generation of neutrino masses of the form $m_{\nu} \sim y_\nu^2 \, v^2 / m_N$, where $v$ is the SM Higgs vev~\cite{Minkowski:1977sc,Yanagida:1979as,Mohapatra:1979ia,GellMann:1980vs,Schechter:1980gr}. However, given the sizes of $\epsilon$ in the viable parameter space of Fig.~\ref{fig:EpsMx}, the resulting neutrino masses are much too small to account for the observed mass splittings in neutrino oscillation experiments~\cite{Gonzalez-Garcia:2014bfa}. Hence, additional physics must be introduced to explain the observed neutrino masses and mixing angles. 

The global $\mathbb{Z}_3$ is a generalization of lepton number in the SM. Hence, if this lepton number is softly broken, for instance, by a small Majorana mass for $N$ or $N^c$, out-of-equilibrium decays of $N$ and $N^c$ may provide a favorable condition for leptogenesis. However, electroweak sphaleron conversion of a lepton asymmetry into a baryon asymmetry demands that $N$ and $N^c$ reheat the SM plasma to a temperature above $\order{100} \GeV$~\cite{Luty:1992un,Fukugita:1986hr}. This restricts $\epsilon \gtrsim \order{10^{-15}} \left( m_{\xp} / 10^{16} \GeV \right)^{-1/2}$. From Fig.~\ref{fig:EpsMx}, this region of parameter space is in mild tension with constraints from neutrino telescopes, although more careful estimates may relax these limits. 

In this Letter, we have considered a new thermal freeze-out mechanism for ultra-heavy DM. Coannihilation with a lighter unstable species exponentially enhances the depletion of the thermal DM number density in the early universe. Despite the fact that the HS must be extremely weakly coupled to the SM, this mechanism motivates interesting experimental signals at neutrino and cosmic ray telescopes. In particular, our model destabilizes any DM candidate, giving rise to phenomenology associated with the late decays of heavy particles. For $\order{1}$ couplings in the HS, this scenario favors DM as heavy as $M_\text{GUT} \sim \order{10^{16}} \GeV$.

\section*{Acknowledgments}

We would like to thank Nikita Blinov, Philip Schuster, Gustavo Marques Tavares, and Natalia Toro for valuable conversations and Gordan Krnjaic for the title recommendation. AB is supported by the U.S. Department of Energy under Contract No. DE-AC02-76SF00515.

\bibliography{coannih}

\begin{thebibliography}{10}%
\makeatletter
\providecommand \@ifxundefined [1]{%
 \ifx #1\undefined \expandafter \@firstoftwo
 \else \expandafter \@secondoftwo
\fi
}%
\providecommand \@ifnum [1]{%
 \ifnum #1\expandafter \@firstoftwo
 \else \expandafter \@secondoftwo
\fi
}%
\providecommand \enquote [1]{``#1''}%
\providecommand \bibnamefont  [1]{#1}%
\providecommand \bibfnamefont [1]{#1}%
\providecommand \citenamefont [1]{#1}%
\providecommand\href[0]{\@sanitize\@href}%
\providecommand\@href[1]{\endgroup\@@startlink{#1}\endgroup\@@href}%
\providecommand\@@href[1]{#1\@@endlink}%
\providecommand \@sanitize [0]{\begingroup\catcode`\&12\catcode`\#12\relax}%
\@ifxundefined \pdfoutput {\@firstoftwo}{%
 \@ifnum{\z@=\pdfoutput}{\@firstoftwo}{\@secondoftwo}%
}{%
 \providecommand\@@startlink[1]{\leavevmode\special{html:<a href="#1">}}%
 \providecommand\@@endlink[0]{\special{html:</a>}}%
}{%
 \providecommand\@@startlink[1]{%
  \leavevmode
  \pdfstartlink
   attr{/Border[0 0 1 ]/H/I/C[0 1 1]}%
   user{/Subtype/Link/A<</Type/Action/S/URI/URI(#1)>>}%
  \relax
 }%
 \providecommand\@@endlink[0]{\pdfendlink}%
}%
\providecommand \url  [0]{\begingroup\@sanitize \@url }%
\providecommand \@url [1]{\endgroup\@href {#1}{\urlprefix}}%
\providecommand \urlprefix [0]{URL }%
\providecommand \Eprint[0]{\href }%
\@ifxundefined \urlstyle {%
  \providecommand \doi [1]{doi:\discretionary{}{}{}#1}%
}{%
  \providecommand \doi [0]{doi:\discretionary{}{}{}\begingroup
  \urlstyle{rm}\Url }%
}%
\providecommand \doibase [0]{http://dx.doi.org/}%
\providecommand \Doi[1]{\href{\doibase#1}}%
\providecommand \bibAnnote [3]{%
  \BibitemShut{#1}%
  \begin{quotation}\noindent
    \textsc{Key:}\ #2\\\textsc{Annotation:}\ #3%
  \end{quotation}%
}%
\providecommand \bibAnnoteFile [2]{%
  \IfFileExists{#2}{\bibAnnote {#1} {#2} {\input{#2}}}{}%
}%
\providecommand \typeout [0]{\immediate \write \m@ne }%
\providecommand \selectlanguage [0]{\@gobble}%
\providecommand \bibinfo [0]{\@secondoftwo}%
\providecommand \bibfield [0]{\@secondoftwo}%
\providecommand \translation [1]{[#1]}%
\providecommand \BibitemOpen[0]{}%
\providecommand \bibitemStop [0]{}%
\providecommand \bibitemNoStop [0]{.\EOS\space}%
\providecommand \EOS [0]{\spacefactor3000\relax}%
\providecommand \BibitemShut [1]{\csname bibitem#1\endcsname}%
\bibitem{Aad:2015zva}%
  \BibitemOpen
  \bibfield{author}{%
  \bibinfo {author} {\bibfnamefont{G.}~\bibnamefont{Aad}} \emph{et~al.}
  (\bibinfo {collaboration} {ATLAS}),\ }%
  \bibfield{journal}{%
  \Doi{10.1140/epjc/s10052-015-3517-3, 10.1140/epjc/s10052-015-3639-7}{\bibinfo
  {journal} {Eur. Phys. J.}}\ }%
  \textbf{\bibinfo {volume} {C75}},\ \bibinfo {pages} {299} (\bibinfo {year}
  {2015}),\ \bibinfo {note} {[Erratum: Eur. Phys. J.C75,no.9,408(2015)]},\
  \Eprint{http://arxiv.org/abs/1502.01518}{arXiv:1502.01518 [hep-ex]}%
  \bibAnnoteFile{NoStop}{Aad:2015zva}%
\bibitem{Khachatryan:2014rra}%
  \BibitemOpen
  \bibfield{author}{%
  \bibinfo {author} {\bibfnamefont{V.}~\bibnamefont{Khachatryan}} \emph{et~al.}
  (\bibinfo {collaboration} {CMS}),\ }%
  \bibfield{journal}{%
  \Doi{10.1140/epjc/s10052-015-3451-4}{\bibinfo {journal} {Eur. Phys. J.}}\ }%
  \textbf{\bibinfo {volume} {C75}},\ \bibinfo {pages} {235} (\bibinfo {year}
  {2015}),\ \Eprint{http://arxiv.org/abs/1408.3583}{arXiv:1408.3583 [hep-ex]}%
  \bibAnnoteFile{NoStop}{Khachatryan:2014rra}%
\bibitem{Tan:2016zwf}%
  \BibitemOpen
  \bibfield{author}{%
  \bibinfo {author} {\bibfnamefont{A.}~\bibnamefont{Tan}} \emph{et~al.}
  (\bibinfo {collaboration} {PandaX-II}),\ }%
  \bibfield{journal}{%
  \Doi{10.1103/PhysRevLett.117.121303}{\bibinfo {journal} {Phys. Rev. Lett.}}\
  }%
  \textbf{\bibinfo {volume} {117}},\ \bibinfo {pages} {121303} (\bibinfo {year}
  {2016}),\ \Eprint{http://arxiv.org/abs/1607.07400}{arXiv:1607.07400
  [hep-ex]}%
  \bibAnnoteFile{NoStop}{Tan:2016zwf}%
\bibitem{Akerib:2015rjg}%
  \BibitemOpen
  \bibfield{author}{%
  \bibinfo {author} {\bibfnamefont{D.~S.}\ \bibnamefont{Akerib}} \emph{et~al.}
  (\bibinfo {collaboration} {LUX}),\ }%
  \bibfield{journal}{%
  \Doi{10.1103/PhysRevLett.116.161301}{\bibinfo {journal} {Phys. Rev. Lett.}}\
  }%
  \textbf{\bibinfo {volume} {116}},\ \bibinfo {pages} {161301} (\bibinfo {year}
  {2016}),\ \Eprint{http://arxiv.org/abs/1512.03506}{arXiv:1512.03506
  [astro-ph.CO]}%
  \bibAnnoteFile{NoStop}{Akerib:2015rjg}%
\bibitem{Akerib:2016vxi}%
  \BibitemOpen
  \bibfield{author}{%
  \bibinfo {author} {\bibfnamefont{D.~S.}\ \bibnamefont{Akerib}} \emph{et~al.}
  (\bibinfo {collaboration} {LUX}),\ }%
  \bibfield{journal}{%
  \Doi{10.1103/PhysRevLett.118.021303}{\bibinfo {journal} {Phys. Rev. Lett.}}\
  }%
  \textbf{\bibinfo {volume} {118}},\ \bibinfo {pages} {021303} (\bibinfo {year}
  {2017}),\ \Eprint{http://arxiv.org/abs/1608.07648}{arXiv:1608.07648
  [astro-ph.CO]}%
  \bibAnnoteFile{NoStop}{Akerib:2016vxi}%
\bibitem{Agnese:2015nto}%
  \BibitemOpen
  \bibfield{author}{%
  \bibinfo {author} {\bibfnamefont{R.}~\bibnamefont{Agnese}} \emph{et~al.}
  (\bibinfo {collaboration} {SuperCDMS}),\ }%
  \bibfield{journal}{%
  \Doi{10.1103/PhysRevLett.116.071301}{\bibinfo {journal} {Phys. Rev. Lett.}}\
  }%
  \textbf{\bibinfo {volume} {116}},\ \bibinfo {pages} {071301} (\bibinfo {year}
  {2016}),\ \Eprint{http://arxiv.org/abs/1509.02448}{arXiv:1509.02448
  [astro-ph.CO]}%
  \bibAnnoteFile{NoStop}{Agnese:2015nto}%
\bibitem{Feng:2008mu}%
  \BibitemOpen
  \bibfield{author}{%
  \bibinfo {author} {\bibfnamefont{J.~L.}\ \bibnamefont{Feng}}, \bibinfo
  {author} {\bibfnamefont{H.}~\bibnamefont{Tu}},\ and\ \bibinfo {author}
  {\bibfnamefont{H.-B.}\ \bibnamefont{Yu}},\ }%
  \bibfield{journal}{%
  \Doi{10.1088/1475-7516/2008/10/043}{\bibinfo {journal} {JCAP}}\ }%
  \textbf{\bibinfo {volume} {0810}},\ \bibinfo {pages} {043} (\bibinfo {year}
  {2008}),\ \Eprint{http://arxiv.org/abs/0808.2318}{arXiv:0808.2318 [hep-ph]}%
  \bibAnnoteFile{NoStop}{Feng:2008mu}%
\bibitem{Hardy:2017wkr}%
  \BibitemOpen
  \bibfield{author}{%
  \bibinfo {author} {\bibfnamefont{E.}~\bibnamefont{Hardy}}\ and\ \bibinfo
  {author} {\bibfnamefont{J.}~\bibnamefont{Unwin}}}%
   (\bibinfo {year} {2017}),\
  \Eprint{http://arxiv.org/abs/1703.07642}{arXiv:1703.07642 [hep-ph]}%
  \bibAnnoteFile{NoStop}{Hardy:2017wkr}%
\bibitem{Adshead:2016xxj}%
  \BibitemOpen
  \bibfield{author}{%
  \bibinfo {author} {\bibfnamefont{P.}~\bibnamefont{Adshead}}, \bibinfo
  {author} {\bibfnamefont{Y.}~\bibnamefont{Cui}},\ and\ \bibinfo {author}
  {\bibfnamefont{J.}~\bibnamefont{Shelton}},\ }%
  \bibfield{journal}{%
  \Doi{10.1007/JHEP06(2016)016}{\bibinfo {journal} {JHEP}}\ }%
  \textbf{\bibinfo {volume} {06}},\ \bibinfo {pages} {016} (\bibinfo {year}
  {2016}),\ \Eprint{http://arxiv.org/abs/1604.02458}{arXiv:1604.02458
  [hep-ph]}%
  \bibAnnoteFile{NoStop}{Adshead:2016xxj}%
\bibitem{Pospelov:2007mp}%
  \BibitemOpen
  \bibfield{author}{%
  \bibinfo {author} {\bibfnamefont{M.}~\bibnamefont{Pospelov}}, \bibinfo
  {author} {\bibfnamefont{A.}~\bibnamefont{Ritz}},\ and\ \bibinfo {author}
  {\bibfnamefont{M.~B.}\ \bibnamefont{Voloshin}},\ }%
  \bibfield{journal}{%
  \Doi{10.1016/j.physletb.2008.02.052}{\bibinfo {journal} {Phys. Lett.}}\ }%
  \textbf{\bibinfo {volume} {B662}},\ \bibinfo {pages} {53} (\bibinfo {year}
  {2008}),\ \Eprint{http://arxiv.org/abs/0711.4866}{arXiv:0711.4866 [hep-ph]}%
  \bibAnnoteFile{NoStop}{Pospelov:2007mp}%
\bibitem{Griest:1989wd}%
  \BibitemOpen
  \bibfield{author}{%
  \bibinfo {author} {\bibfnamefont{K.}~\bibnamefont{Griest}}\ and\ \bibinfo
  {author} {\bibfnamefont{M.}~\bibnamefont{Kamionkowski}},\ }%
  \bibfield{journal}{%
  \Doi{10.1103/PhysRevLett.64.615}{\bibinfo {journal} {Phys. Rev. Lett.}}\ }%
  \textbf{\bibinfo {volume} {64}},\ \bibinfo {pages} {615} (\bibinfo {year}
  {1990})%
  \bibAnnoteFile{NoStop}{Griest:1989wd}%
\bibitem{Fornengo:2002db}%
  \BibitemOpen
  \bibfield{author}{%
  \bibinfo {author} {\bibfnamefont{N.}~\bibnamefont{Fornengo}}, \bibinfo
  {author} {\bibfnamefont{A.}~\bibnamefont{Riotto}},\ and\ \bibinfo {author}
  {\bibfnamefont{S.}~\bibnamefont{Scopel}},\ }%
  \bibfield{journal}{%
  \Doi{10.1103/PhysRevD.67.023514}{\bibinfo {journal} {Phys. Rev.}}\ }%
  \textbf{\bibinfo {volume} {D67}},\ \bibinfo {pages} {023514} (\bibinfo {year}
  {2003}),\ \Eprint{http://arxiv.org/abs/hep-ph/0208072}{arXiv:hep-ph/0208072
  [hep-ph]}%
  \bibAnnoteFile{NoStop}{Fornengo:2002db}%
\bibitem{Kane:2015jia}%
  \BibitemOpen
  \bibfield{author}{%
  \bibinfo {author} {\bibfnamefont{G.}~\bibnamefont{Kane}}, \bibinfo {author}
  {\bibfnamefont{K.}~\bibnamefont{Sinha}},\ and\ \bibinfo {author}
  {\bibfnamefont{S.}~\bibnamefont{Watson}},\ }%
  \bibfield{journal}{%
  \Doi{10.1142/S0218271815300220}{\bibinfo {journal} {Int. J. Mod. Phys.}}\ }%
  \textbf{\bibinfo {volume} {D24}},\ \bibinfo {pages} {1530022} (\bibinfo
  {year} {2015}),\ \Eprint{http://arxiv.org/abs/1502.07746}{arXiv:1502.07746
  [hep-th]}%
  \bibAnnoteFile{NoStop}{Kane:2015jia}%
\bibitem{Hooper:2013nia}%
  \BibitemOpen
  \bibfield{author}{%
  \bibinfo {author} {\bibfnamefont{D.}~\bibnamefont{Hooper}},\ }%
  \bibfield{journal}{%
  \Doi{10.1103/PhysRevD.88.083519}{\bibinfo {journal} {Phys. Rev.}}\ }%
  \textbf{\bibinfo {volume} {D88}},\ \bibinfo {pages} {083519} (\bibinfo {year}
  {2013}),\ \Eprint{http://arxiv.org/abs/1307.0826}{arXiv:1307.0826 [hep-ph]}%
  \bibAnnoteFile{NoStop}{Hooper:2013nia}%
\bibitem{Randall:2015xza}%
  \BibitemOpen
  \bibfield{author}{%
  \bibinfo {author} {\bibfnamefont{L.}~\bibnamefont{Randall}}, \bibinfo
  {author} {\bibfnamefont{J.}~\bibnamefont{Scholtz}},\ and\ \bibinfo {author}
  {\bibfnamefont{J.}~\bibnamefont{Unwin}},\ }%
  \bibfield{journal}{%
  \Doi{10.1007/JHEP03(2016)011}{\bibinfo {journal} {JHEP}}\ }%
  \textbf{\bibinfo {volume} {03}},\ \bibinfo {pages} {011} (\bibinfo {year}
  {2016}),\ \Eprint{http://arxiv.org/abs/1509.08477}{arXiv:1509.08477
  [hep-ph]}%
  \bibAnnoteFile{NoStop}{Randall:2015xza}%
\bibitem{Reece:2015lch}%
  \BibitemOpen
  \bibfield{author}{%
  \bibinfo {author} {\bibfnamefont{M.}~\bibnamefont{Reece}}\ and\ \bibinfo
  {author} {\bibfnamefont{T.}~\bibnamefont{Roxlo}},\ }%
  \bibfield{journal}{%
  \Doi{10.1007/JHEP09(2016)096}{\bibinfo {journal} {JHEP}}\ }%
  \textbf{\bibinfo {volume} {09}},\ \bibinfo {pages} {096} (\bibinfo {year}
  {2016}),\ \Eprint{http://arxiv.org/abs/1511.06768}{arXiv:1511.06768
  [hep-ph]}%
  \bibAnnoteFile{NoStop}{Reece:2015lch}%
\bibitem{Lyth:1995ka}%
  \BibitemOpen
  \bibfield{author}{%
  \bibinfo {author} {\bibfnamefont{D.~H.}\ \bibnamefont{Lyth}}\ and\ \bibinfo
  {author} {\bibfnamefont{E.~D.}\ \bibnamefont{Stewart}},\ }%
  \bibfield{journal}{%
  \Doi{10.1103/PhysRevD.53.1784}{\bibinfo {journal} {Phys. Rev.}}\ }%
  \textbf{\bibinfo {volume} {D53}},\ \bibinfo {pages} {1784} (\bibinfo {year}
  {1996}),\ \Eprint{http://arxiv.org/abs/hep-ph/9510204}{arXiv:hep-ph/9510204
  [hep-ph]}%
  \bibAnnoteFile{NoStop}{Lyth:1995ka}%
\bibitem{Davoudiasl:2015vba}%
  \BibitemOpen
  \bibfield{author}{%
  \bibinfo {author} {\bibfnamefont{H.}~\bibnamefont{Davoudiasl}}, \bibinfo
  {author} {\bibfnamefont{D.}~\bibnamefont{Hooper}},\ and\ \bibinfo {author}
  {\bibfnamefont{S.~D.}\ \bibnamefont{McDermott}},\ }%
  \bibfield{journal}{%
  \Doi{10.1103/PhysRevLett.116.031303}{\bibinfo {journal} {Phys. Rev. Lett.}}\
  }%
  \textbf{\bibinfo {volume} {116}},\ \bibinfo {pages} {031303} (\bibinfo {year}
  {2016}),\ \Eprint{http://arxiv.org/abs/1507.08660}{arXiv:1507.08660
  [hep-ph]}%
  \bibAnnoteFile{NoStop}{Davoudiasl:2015vba}%
\bibitem{Cohen:2008nb}%
  \BibitemOpen
  \bibfield{author}{%
  \bibinfo {author} {\bibfnamefont{T.}~\bibnamefont{Cohen}}, \bibinfo {author}
  {\bibfnamefont{D.~E.}\ \bibnamefont{Morrissey}},\ and\ \bibinfo {author}
  {\bibfnamefont{A.}~\bibnamefont{Pierce}},\ }%
  \bibfield{journal}{%
  \Doi{10.1103/PhysRevD.78.111701}{\bibinfo {journal} {Phys. Rev.}}\ }%
  \textbf{\bibinfo {volume} {D78}},\ \bibinfo {pages} {111701} (\bibinfo {year}
  {2008}),\ \Eprint{http://arxiv.org/abs/0808.3994}{arXiv:0808.3994 [hep-ph]}%
  \bibAnnoteFile{NoStop}{Cohen:2008nb}%
\bibitem{Berlin:2016vnh}%
  \BibitemOpen
  \bibfield{author}{%
  \bibinfo {author} {\bibfnamefont{A.}~\bibnamefont{Berlin}}, \bibinfo {author}
  {\bibfnamefont{D.}~\bibnamefont{Hooper}},\ and\ \bibinfo {author}
  {\bibfnamefont{G.}~\bibnamefont{Krnjaic}},\ }%
  \bibfield{journal}{%
  \Doi{10.1016/j.physletb.2016.06.037}{\bibinfo {journal} {Phys. Lett.}}\ }%
  \textbf{\bibinfo {volume} {B760}},\ \bibinfo {pages} {106} (\bibinfo {year}
  {2016}),\ \Eprint{http://arxiv.org/abs/1602.08490}{arXiv:1602.08490
  [hep-ph]}%
  \bibAnnoteFile{NoStop}{Berlin:2016vnh}%
\bibitem{Berlin:2016gtr}%
  \BibitemOpen
  \bibfield{author}{%
  \bibinfo {author} {\bibfnamefont{A.}~\bibnamefont{Berlin}}, \bibinfo {author}
  {\bibfnamefont{D.}~\bibnamefont{Hooper}},\ and\ \bibinfo {author}
  {\bibfnamefont{G.}~\bibnamefont{Krnjaic}},\ }%
  \bibfield{journal}{%
  \Doi{10.1103/PhysRevD.94.095019}{\bibinfo {journal} {Phys. Rev.}}\ }%
  \textbf{\bibinfo {volume} {D94}},\ \bibinfo {pages} {095019} (\bibinfo {year}
  {2016}),\ \Eprint{http://arxiv.org/abs/1609.02555}{arXiv:1609.02555
  [hep-ph]}%
  \bibAnnoteFile{NoStop}{Berlin:2016gtr}%
\bibitem{Kolb:1998ki}%
  \BibitemOpen
  \bibfield{author}{%
  \bibinfo {author} {\bibfnamefont{E.~W.}\ \bibnamefont{Kolb}}, \bibinfo
  {author} {\bibfnamefont{D.~J.~H.}\ \bibnamefont{Chung}},\ and\ \bibinfo
  {author} {\bibfnamefont{A.}~\bibnamefont{Riotto}},\ }%
  in\ \emph{\bibinfo {booktitle} {{Trends in theoretical physics II.
  Proceedings, 2nd La Plata Meeting, Buenos Aires, Argentina, November
  29-December 4, 1998}}}\ (\bibinfo {year} {1998})\ pp.\ \bibinfo {pages}
  {91--105},\ \bibinfo {note} {[,91(1998)]},\
  \Eprint{http://arxiv.org/abs/hep-ph/9810361}{arXiv:hep-ph/9810361 [hep-ph]},\
  \url{http://lss.fnal.gov/cgi-bin/find_paper.pl?conf-98-325}%
  \bibAnnoteFile{NoStop}{Kolb:1998ki}%
\bibitem{Chung:1998rq}%
  \BibitemOpen
  \bibfield{author}{%
  \bibinfo {author} {\bibfnamefont{D.~J.~H.}\ \bibnamefont{Chung}}, \bibinfo
  {author} {\bibfnamefont{E.~W.}\ \bibnamefont{Kolb}},\ and\ \bibinfo {author}
  {\bibfnamefont{A.}~\bibnamefont{Riotto}},\ }%
  \bibfield{journal}{%
  \Doi{10.1103/PhysRevD.60.063504}{\bibinfo {journal} {Phys. Rev.}}\ }%
  \textbf{\bibinfo {volume} {D60}},\ \bibinfo {pages} {063504} (\bibinfo {year}
  {1999}),\ \Eprint{http://arxiv.org/abs/hep-ph/9809453}{arXiv:hep-ph/9809453
  [hep-ph]}%
  \bibAnnoteFile{NoStop}{Chung:1998rq}%
\bibitem{Chung:2001cb}%
  \BibitemOpen
  \bibfield{author}{%
  \bibinfo {author} {\bibfnamefont{D.~J.~H.}\ \bibnamefont{Chung}}, \bibinfo
  {author} {\bibfnamefont{P.}~\bibnamefont{Crotty}}, \bibinfo {author}
  {\bibfnamefont{E.~W.}\ \bibnamefont{Kolb}},\ and\ \bibinfo {author}
  {\bibfnamefont{A.}~\bibnamefont{Riotto}},\ }%
  \bibfield{journal}{%
  \Doi{10.1103/PhysRevD.64.043503}{\bibinfo {journal} {Phys. Rev.}}\ }%
  \textbf{\bibinfo {volume} {D64}},\ \bibinfo {pages} {043503} (\bibinfo {year}
  {2001}),\ \Eprint{http://arxiv.org/abs/hep-ph/0104100}{arXiv:hep-ph/0104100
  [hep-ph]}%
  \bibAnnoteFile{NoStop}{Chung:2001cb}%
\bibitem{Feldstein:2013uha}%
  \BibitemOpen
  \bibfield{author}{%
  \bibinfo {author} {\bibfnamefont{B.}~\bibnamefont{Feldstein}}, \bibinfo
  {author} {\bibfnamefont{M.}~\bibnamefont{Ibe}},\ and\ \bibinfo {author}
  {\bibfnamefont{T.~T.}\ \bibnamefont{Yanagida}},\ }%
  \bibfield{journal}{%
  \Doi{10.1103/PhysRevLett.112.101301}{\bibinfo {journal} {Phys. Rev. Lett.}}\
  }%
  \textbf{\bibinfo {volume} {112}},\ \bibinfo {pages} {101301} (\bibinfo {year}
  {2014}),\ \Eprint{http://arxiv.org/abs/1310.7495}{arXiv:1310.7495 [hep-ph]}%
  \bibAnnoteFile{NoStop}{Feldstein:2013uha}%
\bibitem{Harigaya:2014waa}%
  \BibitemOpen
  \bibfield{author}{%
  \bibinfo {author} {\bibfnamefont{K.}~\bibnamefont{Harigaya}}, \bibinfo
  {author} {\bibfnamefont{M.}~\bibnamefont{Kawasaki}}, \bibinfo {author}
  {\bibfnamefont{K.}~\bibnamefont{Mukaida}},\ and\ \bibinfo {author}
  {\bibfnamefont{M.}~\bibnamefont{Yamada}},\ }%
  \bibfield{journal}{%
  \Doi{10.1103/PhysRevD.89.083532}{\bibinfo {journal} {Phys. Rev.}}\ }%
  \textbf{\bibinfo {volume} {D89}},\ \bibinfo {pages} {083532} (\bibinfo {year}
  {2014}),\ \Eprint{http://arxiv.org/abs/1402.2846}{arXiv:1402.2846 [hep-ph]}%
  \bibAnnoteFile{NoStop}{Harigaya:2014waa}%
\bibitem{Hui:1998dc}%
  \BibitemOpen
  \bibfield{author}{%
  \bibinfo {author} {\bibfnamefont{L.}~\bibnamefont{Hui}}\ and\ \bibinfo
  {author} {\bibfnamefont{E.~D.}\ \bibnamefont{Stewart}},\ }%
  \bibfield{journal}{%
  \Doi{10.1103/PhysRevD.60.023518}{\bibinfo {journal} {Phys. Rev.}}\ }%
  \textbf{\bibinfo {volume} {D60}},\ \bibinfo {pages} {023518} (\bibinfo {year}
  {1999}),\ \Eprint{http://arxiv.org/abs/hep-ph/9812345}{arXiv:hep-ph/9812345
  [hep-ph]}%
  \bibAnnoteFile{NoStop}{Hui:1998dc}%
\bibitem{Bramante:2017obj}%
  \BibitemOpen
  \bibfield{author}{%
  \bibinfo {author} {\bibfnamefont{J.}~\bibnamefont{Bramante}}\ and\ \bibinfo
  {author} {\bibfnamefont{J.}~\bibnamefont{Unwin}},\ }%
  \bibfield{journal}{%
  \Doi{10.1007/JHEP02(2017)119}{\bibinfo {journal} {JHEP}}\ }%
  \textbf{\bibinfo {volume} {02}},\ \bibinfo {pages} {119} (\bibinfo {year}
  {2017}),\ \Eprint{http://arxiv.org/abs/1701.05859}{arXiv:1701.05859
  [hep-ph]}%
  \bibAnnoteFile{NoStop}{Bramante:2017obj}%
\bibitem{Harigaya:2016vda}%
  \BibitemOpen
  \bibfield{author}{%
  \bibinfo {author} {\bibfnamefont{K.}~\bibnamefont{Harigaya}}, \bibinfo
  {author} {\bibfnamefont{T.}~\bibnamefont{Lin}},\ and\ \bibinfo {author}
  {\bibfnamefont{H.~K.}\ \bibnamefont{Lou}},\ }%
  \bibfield{journal}{%
  \Doi{10.1007/JHEP09(2016)014}{\bibinfo {journal} {JHEP}}\ }%
  \textbf{\bibinfo {volume} {09}},\ \bibinfo {pages} {014} (\bibinfo {year}
  {2016}),\ \Eprint{http://arxiv.org/abs/1606.00923}{arXiv:1606.00923
  [hep-ph]}%
  \bibAnnoteFile{NoStop}{Harigaya:2016vda}%
\bibitem{Griest:1990kh}%
  \BibitemOpen
  \bibfield{author}{%
  \bibinfo {author} {\bibfnamefont{K.}~\bibnamefont{Griest}}\ and\ \bibinfo
  {author} {\bibfnamefont{D.}~\bibnamefont{Seckel}},\ }%
  \bibfield{journal}{%
  \Doi{10.1103/PhysRevD.43.3191}{\bibinfo {journal} {Phys. Rev.}}\ }%
  \textbf{\bibinfo {volume} {D43}},\ \bibinfo {pages} {3191} (\bibinfo {year}
  {1991})%
  \bibAnnoteFile{NoStop}{Griest:1990kh}%
\bibitem{Edsjo:1997bg}%
  \BibitemOpen
  \bibfield{author}{%
  \bibinfo {author} {\bibfnamefont{J.}~\bibnamefont{Edsjo}}\ and\ \bibinfo
  {author} {\bibfnamefont{P.}~\bibnamefont{Gondolo}},\ }%
  \bibfield{journal}{%
  \Doi{10.1103/PhysRevD.56.1879}{\bibinfo {journal} {Phys. Rev.}}\ }%
  \textbf{\bibinfo {volume} {D56}},\ \bibinfo {pages} {1879} (\bibinfo {year}
  {1997}),\ \Eprint{http://arxiv.org/abs/hep-ph/9704361}{arXiv:hep-ph/9704361
  [hep-ph]}%
  \bibAnnoteFile{NoStop}{Edsjo:1997bg}%
\bibitem{Cline:2017tka}%
  \BibitemOpen
  \bibfield{author}{%
  \bibinfo {author} {\bibfnamefont{J.}~\bibnamefont{Cline}}, \bibinfo {author}
  {\bibfnamefont{H.}~\bibnamefont{Liu}}, \bibinfo {author}
  {\bibfnamefont{T.}~\bibnamefont{Slatyer}},\ and\ \bibinfo {author}
  {\bibfnamefont{W.}~\bibnamefont{Xue}}}%
   (\bibinfo {year} {2017}),\
  \Eprint{http://arxiv.org/abs/1702.07716}{arXiv:1702.07716 [hep-ph]}%
  \bibAnnoteFile{NoStop}{Cline:2017tka}%
\bibitem{Dey:2016qgf}%
  \BibitemOpen
  \bibfield{author}{%
  \bibinfo {author} {\bibfnamefont{U.~K.}\ \bibnamefont{Dey}}, \bibinfo
  {author} {\bibfnamefont{T.~N.}\ \bibnamefont{Maity}},\ and\ \bibinfo {author}
  {\bibfnamefont{T.~S.}\ \bibnamefont{Ray}},\ }%
  \bibfield{journal}{%
  \Doi{10.1088/1475-7516/2017/03/045}{\bibinfo {journal} {JCAP}}\ }%
  \textbf{\bibinfo {volume} {1703}},\ \bibinfo {pages} {045} (\bibinfo {year}
  {2017}),\ \Eprint{http://arxiv.org/abs/1612.09074}{arXiv:1612.09074
  [hep-ph]}%
  \bibAnnoteFile{NoStop}{Dey:2016qgf}%
\bibitem{Allahverdi:2010xz}%
  \BibitemOpen
  \bibfield{author}{%
  \bibinfo {author} {\bibfnamefont{R.}~\bibnamefont{Allahverdi}}, \bibinfo
  {author} {\bibfnamefont{R.}~\bibnamefont{Brandenberger}}, \bibinfo {author}
  {\bibfnamefont{F.-Y.}\ \bibnamefont{Cyr-Racine}},\ and\ \bibinfo {author}
  {\bibfnamefont{A.}~\bibnamefont{Mazumdar}},\ }%
  \bibfield{journal}{%
  \Doi{10.1146/annurev.nucl.012809.104511}{\bibinfo {journal} {Ann. Rev. Nucl.
  Part. Sci.}}\ }%
  \textbf{\bibinfo {volume} {60}},\ \bibinfo {pages} {27} (\bibinfo {year}
  {2010}),\ \Eprint{http://arxiv.org/abs/1001.2600}{arXiv:1001.2600 [hep-th]}%
  \bibAnnoteFile{NoStop}{Allahverdi:2010xz}%
\bibitem{Poulin:2016nat}%
  \BibitemOpen
  \bibfield{author}{%
  \bibinfo {author} {\bibfnamefont{V.}~\bibnamefont{Poulin}}, \bibinfo {author}
  {\bibfnamefont{P.~D.}\ \bibnamefont{Serpico}},\ and\ \bibinfo {author}
  {\bibfnamefont{J.}~\bibnamefont{Lesgourgues}},\ }%
  \bibfield{journal}{%
  \Doi{10.1088/1475-7516/2016/08/036}{\bibinfo {journal} {JCAP}}\ }%
  \textbf{\bibinfo {volume} {1608}},\ \bibinfo {pages} {036} (\bibinfo {year}
  {2016}),\ \Eprint{http://arxiv.org/abs/1606.02073}{arXiv:1606.02073
  [astro-ph.CO]}%
  \bibAnnoteFile{NoStop}{Poulin:2016nat}%
\bibitem{Kolb:1990vq}%
  \BibitemOpen
  \bibfield{author}{%
  \bibinfo {author} {\bibfnamefont{E.~W.}\ \bibnamefont{Kolb}}\ and\ \bibinfo
  {author} {\bibfnamefont{M.~S.}\ \bibnamefont{Turner}},\ }%
  \bibfield{journal}{%
  \bibinfo {journal} {Front. Phys.}\ }%
  \textbf{\bibinfo {volume} {69}},\ \bibinfo {pages} {1} (\bibinfo {year}
  {1990})%
  \bibAnnoteFile{NoStop}{Kolb:1990vq}%
\bibitem{Esmaili:2012us}%
  \BibitemOpen
  \bibfield{author}{%
  \bibinfo {author} {\bibfnamefont{A.}~\bibnamefont{Esmaili}}, \bibinfo
  {author} {\bibfnamefont{A.}~\bibnamefont{Ibarra}},\ and\ \bibinfo {author}
  {\bibfnamefont{O.~L.~G.}\ \bibnamefont{Peres}},\ }%
  \bibfield{journal}{%
  \Doi{10.1088/1475-7516/2012/11/034}{\bibinfo {journal} {JCAP}}\ }%
  \textbf{\bibinfo {volume} {1211}},\ \bibinfo {pages} {034} (\bibinfo {year}
  {2012}),\ \Eprint{http://arxiv.org/abs/1205.5281}{arXiv:1205.5281 [hep-ph]}%
  \bibAnnoteFile{NoStop}{Esmaili:2012us}%
\bibitem{Kuznetsov:2016fjt}%
  \BibitemOpen
  \bibfield{author}{%
  \bibinfo {author} {\bibfnamefont{M.~{\relax Yu}.}\ \bibnamefont{Kuznetsov}}}%
   (\bibinfo {year} {2016}),\
  \Eprint{http://arxiv.org/abs/1611.08684}{arXiv:1611.08684 [astro-ph.HE]}%
  \bibAnnoteFile{NoStop}{Kuznetsov:2016fjt}%
\bibitem{Cohen:2016uyg}%
  \BibitemOpen
  \bibfield{author}{%
  \bibinfo {author} {\bibfnamefont{T.}~\bibnamefont{Cohen}}, \bibinfo {author}
  {\bibfnamefont{K.}~\bibnamefont{Murase}}, \bibinfo {author}
  {\bibfnamefont{N.~L.}\ \bibnamefont{Rodd}}, \bibinfo {author}
  {\bibfnamefont{B.~R.}\ \bibnamefont{Safdi}},\ and\ \bibinfo {author}
  {\bibfnamefont{Y.}~\bibnamefont{Soreq}}}%
   (\bibinfo {year} {2016}),\
  \Eprint{http://arxiv.org/abs/1612.05638}{arXiv:1612.05638 [hep-ph]}%
  \bibAnnoteFile{NoStop}{Cohen:2016uyg}%
\bibitem{Martin:2010kz}%
  \BibitemOpen
  \bibfield{author}{%
  \bibinfo {author} {\bibfnamefont{J.}~\bibnamefont{Martin}}\ and\ \bibinfo
  {author} {\bibfnamefont{C.}~\bibnamefont{Ringeval}},\ }%
  \bibfield{journal}{%
  \Doi{10.1103/PhysRevD.82.023511}{\bibinfo {journal} {Phys. Rev.}}\ }%
  \textbf{\bibinfo {volume} {D82}},\ \bibinfo {pages} {023511} (\bibinfo {year}
  {2010}),\ \Eprint{http://arxiv.org/abs/1004.5525}{arXiv:1004.5525
  [astro-ph.CO]}%
  \bibAnnoteFile{NoStop}{Martin:2010kz}%
\bibitem{Adshead:2010mc}%
  \BibitemOpen
  \bibfield{author}{%
  \bibinfo {author} {\bibfnamefont{P.}~\bibnamefont{Adshead}}, \bibinfo
  {author} {\bibfnamefont{R.}~\bibnamefont{Easther}}, \bibinfo {author}
  {\bibfnamefont{J.}~\bibnamefont{Pritchard}},\ and\ \bibinfo {author}
  {\bibfnamefont{A.}~\bibnamefont{Loeb}},\ }%
  \bibfield{journal}{%
  \Doi{10.1088/1475-7516/2011/02/021}{\bibinfo {journal} {JCAP}}\ }%
  \textbf{\bibinfo {volume} {1102}},\ \bibinfo {pages} {021} (\bibinfo {year}
  {2011}),\ \Eprint{http://arxiv.org/abs/1007.3748}{arXiv:1007.3748
  [astro-ph.CO]}%
  \bibAnnoteFile{NoStop}{Adshead:2010mc}%
\bibitem{Martin:2014nya}%
  \BibitemOpen
  \bibfield{author}{%
  \bibinfo {author} {\bibfnamefont{J.}~\bibnamefont{Martin}}, \bibinfo {author}
  {\bibfnamefont{C.}~\bibnamefont{Ringeval}},\ and\ \bibinfo {author}
  {\bibfnamefont{V.}~\bibnamefont{Vennin}},\ }%
  \bibfield{journal}{%
  \Doi{10.1103/PhysRevLett.114.081303}{\bibinfo {journal} {Phys. Rev. Lett.}}\
  }%
  \textbf{\bibinfo {volume} {114}},\ \bibinfo {pages} {081303} (\bibinfo {year}
  {2015}),\ \Eprint{http://arxiv.org/abs/1410.7958}{arXiv:1410.7958
  [astro-ph.CO]}%
  \bibAnnoteFile{NoStop}{Martin:2014nya}%
\bibitem{Dai:2014jja}%
  \BibitemOpen
  \bibfield{author}{%
  \bibinfo {author} {\bibfnamefont{L.}~\bibnamefont{Dai}}, \bibinfo {author}
  {\bibfnamefont{M.}~\bibnamefont{Kamionkowski}},\ and\ \bibinfo {author}
  {\bibfnamefont{J.}~\bibnamefont{Wang}},\ }%
  \bibfield{journal}{%
  \Doi{10.1103/PhysRevLett.113.041302}{\bibinfo {journal} {Phys. Rev. Lett.}}\
  }%
  \textbf{\bibinfo {volume} {113}},\ \bibinfo {pages} {041302} (\bibinfo {year}
  {2014}),\ \Eprint{http://arxiv.org/abs/1404.6704}{arXiv:1404.6704
  [astro-ph.CO]}%
  \bibAnnoteFile{NoStop}{Dai:2014jja}%
\bibitem{Cook:2015vqa}%
  \BibitemOpen
  \bibfield{author}{%
  \bibinfo {author} {\bibfnamefont{J.~L.}\ \bibnamefont{Cook}}, \bibinfo
  {author} {\bibfnamefont{E.}~\bibnamefont{Dimastrogiovanni}}, \bibinfo
  {author} {\bibfnamefont{D.~A.}\ \bibnamefont{Easson}},\ and\ \bibinfo
  {author} {\bibfnamefont{L.~M.}\ \bibnamefont{Krauss}},\ }%
  \bibfield{journal}{%
  \Doi{10.1088/1475-7516/2015/04/047}{\bibinfo {journal} {JCAP}}\ }%
  \textbf{\bibinfo {volume} {1504}},\ \bibinfo {pages} {047} (\bibinfo {year}
  {2015}),\ \Eprint{http://arxiv.org/abs/1502.04673}{arXiv:1502.04673
  [astro-ph.CO]}%
  \bibAnnoteFile{NoStop}{Cook:2015vqa}%
\bibitem{Domcke:2015iaa}%
  \BibitemOpen
  \bibfield{author}{%
  \bibinfo {author} {\bibfnamefont{V.}~\bibnamefont{Domcke}}\ and\ \bibinfo
  {author} {\bibfnamefont{J.}~\bibnamefont{Heisig}},\ }%
  \bibfield{journal}{%
  \Doi{10.1103/PhysRevD.92.103515}{\bibinfo {journal} {Phys. Rev.}}\ }%
  \textbf{\bibinfo {volume} {D92}},\ \bibinfo {pages} {103515} (\bibinfo {year}
  {2015}),\ \Eprint{http://arxiv.org/abs/1504.00345}{arXiv:1504.00345
  [astro-ph.CO]}%
  \bibAnnoteFile{NoStop}{Domcke:2015iaa}%
\bibitem{Minkowski:1977sc}%
  \BibitemOpen
  \bibfield{author}{%
  \bibinfo {author} {\bibfnamefont{P.}~\bibnamefont{Minkowski}},\ }%
  \bibfield{journal}{%
  \Doi{10.1016/0370-2693(77)90435-X}{\bibinfo {journal} {Phys. Lett.}}\ }%
  \textbf{\bibinfo {volume} {B67}},\ \bibinfo {pages} {421} (\bibinfo {year}
  {1977})%
  \bibAnnoteFile{NoStop}{Minkowski:1977sc}%
\bibitem{Yanagida:1979as}%
  \BibitemOpen
  \bibfield{author}{%
  \bibinfo {author} {\bibfnamefont{T.}~\bibnamefont{Yanagida}},\ }%
  \bibfield{booktitle}{%
  \emph{\bibinfo {booktitle} {{Proceedings: Workshop on the Unified Theories
  and the Baryon Number in the Universe: Tsukuba, Japan, February 13-14,
  1979}}},\ }%
  \bibfield{journal}{%
  \bibinfo {journal} {Conf. Proc.}\ }%
  \textbf{\bibinfo {volume} {C7902131}},\ \bibinfo {pages} {95} (\bibinfo
  {year} {1979})%
  \bibAnnoteFile{NoStop}{Yanagida:1979as}%
\bibitem{Mohapatra:1979ia}%
  \BibitemOpen
  \bibfield{author}{%
  \bibinfo {author} {\bibfnamefont{R.~N.}\ \bibnamefont{Mohapatra}}\ and\
  \bibinfo {author} {\bibfnamefont{G.}~\bibnamefont{Senjanovic}},\ }%
  \bibfield{journal}{%
  \Doi{10.1103/PhysRevLett.44.912}{\bibinfo {journal} {Phys. Rev. Lett.}}\ }%
  \textbf{\bibinfo {volume} {44}},\ \bibinfo {pages} {912} (\bibinfo {year}
  {1980})%
  \bibAnnoteFile{NoStop}{Mohapatra:1979ia}%
\bibitem{GellMann:1980vs}%
  \BibitemOpen
  \bibfield{author}{%
  \bibinfo {author} {\bibfnamefont{M.}~\bibnamefont{Gell-Mann}}, \bibinfo
  {author} {\bibfnamefont{P.}~\bibnamefont{Ramond}},\ and\ \bibinfo {author}
  {\bibfnamefont{R.}~\bibnamefont{Slansky}},\ }%
  \bibfield{booktitle}{%
  \emph{\bibinfo {booktitle} {{Supergravity Workshop Stony Brook, New York,
  September 27-28, 1979}}},\ }%
  \bibfield{journal}{%
  \bibinfo {journal} {Conf. Proc.}\ }%
  \textbf{\bibinfo {volume} {C790927}},\ \bibinfo {pages} {315} (\bibinfo
  {year} {1979}),\ \Eprint{http://arxiv.org/abs/1306.4669}{arXiv:1306.4669
  [hep-th]}%
  \bibAnnoteFile{NoStop}{GellMann:1980vs}%
\bibitem{Schechter:1980gr}%
  \BibitemOpen
  \bibfield{author}{%
  \bibinfo {author} {\bibfnamefont{J.}~\bibnamefont{Schechter}}\ and\ \bibinfo
  {author} {\bibfnamefont{J.~W.~F.}\ \bibnamefont{Valle}},\ }%
  \bibfield{journal}{%
  \Doi{10.1103/PhysRevD.22.2227}{\bibinfo {journal} {Phys. Rev.}}\ }%
  \textbf{\bibinfo {volume} {D22}},\ \bibinfo {pages} {2227} (\bibinfo {year}
  {1980})%
  \bibAnnoteFile{NoStop}{Schechter:1980gr}%
\bibitem{Gonzalez-Garcia:2014bfa}%
  \BibitemOpen
  \bibfield{author}{%
  \bibinfo {author} {\bibfnamefont{M.~C.}\ \bibnamefont{Gonzalez-Garcia}},
  \bibinfo {author} {\bibfnamefont{M.}~\bibnamefont{Maltoni}},\ and\ \bibinfo
  {author} {\bibfnamefont{T.}~\bibnamefont{Schwetz}},\ }%
  \bibfield{journal}{%
  \Doi{10.1007/JHEP11(2014)052}{\bibinfo {journal} {JHEP}}\ }%
  \textbf{\bibinfo {volume} {11}},\ \bibinfo {pages} {052} (\bibinfo {year}
  {2014}),\ \Eprint{http://arxiv.org/abs/1409.5439}{arXiv:1409.5439 [hep-ph]}%
  \bibAnnoteFile{NoStop}{Gonzalez-Garcia:2014bfa}%
\bibitem{Luty:1992un}%
  \BibitemOpen
  \bibfield{author}{%
  \bibinfo {author} {\bibfnamefont{M.~A.}\ \bibnamefont{Luty}},\ }%
  \bibfield{journal}{%
  \Doi{10.1103/PhysRevD.45.455}{\bibinfo {journal} {Phys. Rev.}}\ }%
  \textbf{\bibinfo {volume} {D45}},\ \bibinfo {pages} {455} (\bibinfo {year}
  {1992})%
  \bibAnnoteFile{NoStop}{Luty:1992un}%
\bibitem{Fukugita:1986hr}%
  \BibitemOpen
  \bibfield{author}{%
  \bibinfo {author} {\bibfnamefont{M.}~\bibnamefont{Fukugita}}\ and\ \bibinfo
  {author} {\bibfnamefont{T.}~\bibnamefont{Yanagida}},\ }%
  \bibfield{journal}{%
  \Doi{10.1016/0370-2693(86)91126-3}{\bibinfo {journal} {Phys. Lett.}}\ }%
  \textbf{\bibinfo {volume} {B174}},\ \bibinfo {pages} {45} (\bibinfo {year}
  {1986})%
  \bibAnnoteFile{NoStop}{Fukugita:1986hr}%
\end{thebibliography}%

\end{document}